\documentclass[final,3p,times,twocolumn,authoryear]{elsarticle}

\usepackage{amssymb}

\usepackage{color,rotating}

\newcommand{\la}{\mathrel{\mathchoice {\vcenter{\offinterlineskip\halign{\hfil
  $\displaystyle##$\hfil\cr<\cr\sim\cr}}}
  {\vcenter{\offinterlineskip\halign{\hfil$\textstyle##$\hfil\cr
  <\cr\sim\cr}}}
  {\vcenter{\offinterlineskip\halign{\hfil$\scriptstyle##$\hfil\cr
  <\cr\sim\cr}}}
  {\vcenter{\offinterlineskip\halign{\hfil$\scriptscriptstyle##$\hfil\cr
  <\cr\sim\cr}}}}}

\journal{Planetary and Space Science}

\begin{document}

\begin{frontmatter}

\title{The accumulation and trapping of grains at planet gaps:\\
effects of grain growth and fragmentation}

\author[a]{J.-F. Gonzalez\corref{cor}}
\cortext[cor]{Corresponding author}
\ead{Jean-Francois.Gonzalez@ens-lyon.fr}
\author[b]{G. Laibe}
\author[c]{S. T. Maddison}
\author[d,e]{C. Pinte}
\author[d,e]{F. M\'enard}

\address[a]{Universit\'e de Lyon, Lyon, F-69003, France; Universit\'e Lyon~1, Observatoire de Lyon, 9 avenue Charles Andr\'e, Saint-Genis Laval, F-69230, France; CNRS, UMR 5574, Centre de Recherche Astrophysique de Lyon; \'Ecole Normale Sup\'erieure de Lyon, Lyon, F-69007, France}
\address[b]{School of Physics and Astronomy, University of Saint Andrews, North Haugh, St Andrews, Fife KY16 9SS, UK}
\address[c]{Centre for Astrophysics and Supercomputing, Swinburne Institute of Technology, PO Box 218, Hawthorn, VIC 3122, Australia}
\address[d]{UMI-FCA, CNRS/INSU France (UMI 3386), and Departamento de Astronom{\'\i}a, Universidad de Chile, Casilla 36-D Santiago, Chile}
\address[e]{UJF-Grenoble 1 / CNRS-INSU, Institut de Plan\'etologie et d'Astrophysique de Grenoble, UMR 5274, Grenoble, F-38041, France}

\begin{abstract}
We model the dust evolution in protoplanetary disks with full 3D, Smoothed Particle Hydrodynamics (SPH), two-phase (gas+dust) hydrodynamical simulations. The gas+dust dynamics, where aerodynamic drag leads to the vertical settling and radial migration of grains, is consistently treated. In a previous work, we characterized the spatial distribution of non-growing dust grains of different sizes in a disk containing a gap-opening planet and investigated the gap's detectability with the Atacama Large Millimeter/submillimeter Array (ALMA). Here we take into account the effects of grain growth and fragmentation and study their impact on the distribution of solids in the disk. We show that rapid grain growth in the two accumulation zones around planet gaps is strongly affected by fragmentation. We discuss the consequences for ALMA observations.
\end{abstract}


\end{frontmatter}

\section{Introduction}
\label{Sec:Intro}

Planets are believed to form from the aggregation of sub-micronic dust particles in the protoplanetary disk surrounding a nascent star. In the core-accretion scenario \citep[e.g.][]{Alibert2005}, small grains grow to form km-sized planetesimals, which in turn become planetary cores of a few terrestrial masses. These cores then capture gas, initially slowly then in a runaway process to form gas giant planets. In the disk instability scenario \cite[see e.g.][]{Boss2011}, the disk fragments to form clumps that gravitationally collapse to directly form gas giants. Rocky planets are thought to be the result of the runaway accretion of planetesimals followed by oligarchic accretion, followed by merging of the largest embryos. Whereas small dust grains easily stick during collisions to form aggregates up to cm or dm sizes, the subsequent growth to planetesimal size is probably the biggest problem in the theory of planet formation. Several processes preventing this step have been identified and are collectively known as the ``barriers'' to planet formation.

The radial-drift barrier \citep{W1977,NSH1986,Birnstiel2010,Laibe2012,Laibe2014} is the oldest and most widely known of these problems. In a gas disk, dust grains are subject to aerodynamic drag caused by the differential velocity between the sub-Keplerian gas (partially supported by its radial pressure gradient) and the Keplerian dust. The ensuing transfer of linear and angular momentum between both phases results in the vertical settling (towards the disk midplane) and radial migration (towards the star) of dust grains. The efficiency of this process is measured by the Stokes number St, defined as the ratio of the stopping time (i.e. the drag timescale) to the Keplerian orbital time. Small grains (with $\mathrm{St}\ll1$) are strongly coupled to the gas and follow its motion, whereas large grains (with $\mathrm{St}\gg1$) are largely insensitive to gas drag and stay on their Keplerian orbits. Dust settling and migration are most efficient for intermediate sizes and in particular, radial drift is fastest when $\mathrm{St}\sim1$. Grains of the corresponding size can thus migrate through the disk and fall onto the star in a very short time compared to the disk lifetime and the planet formation timescale, both of a few million years. This ``optimal'' size for migration and the infall timescale both depend on the nebular conditions and the distance to the central star, and  typically range from centimeter to meter sizes and from hundreds to tens of thousands of years, respectively. Grains growing from sub-$\mu$m size and reaching $\mathrm{St}=1$ should therefore be lost from the disk before forming planetesimals: this is the ``radial-drift barrier''. However, \citet{Laibe2008} found that migrating, settling and growing, but non-fragmenting, grains can overcome $\mathrm{St}=1$ and decouple from the gas in the inner regions of typical disks, where they can continue to grow without migrating. More recently \citet{Laibe2012}, confirming and extending a previous work by \citet{YS2002}, showed that even non-growing grains can slow their migration and pile up in the inner regions of disks with steep surface density profiles and shallow temperature profiles, consistent with the majority of observed protoplanetary disks. This is because, under these specific disk conditions, for inwards migrating grains gas drag increases faster than the acceleration due to the radial pressure gradient. \citet{Laibe2014} showed that in typical disks, the interplay between growth (again without fragmentation) and migration dramatically amplifies this pile-up and literally stops the grains, corroborating the results found numerically in \citet{Laibe2008}, and giving a potential solution to the radial-drift barrier problem, which depends only on the local temperature and pressure conditions in the disk.

The second barrier to planet formation comes from the collisions of dust grains. The relative velocities between grains increase with their size \citep{WC1993}: whereas small grains typically collide at a few cm\,s$^{-1}$ and can stick very efficiently thanks to van der Waals forces, particles larger than cm sizes have relative velocities exceeding a m\,s $^{-1}$. Their relative kinetic energy causes particles to break upon impact and produce a number of smaller fragments, therefore preventing their growth to larger sizes: this is called the ``fragmentation barrier'' \citep{DD2005,Blum2008}. 
More recently, a third barrier has been identified: millimeter grains can bounce upon collision at velocities lower than the fragmentation threshold \citep{Zsom2010,Windmark2012}. This also prevents further growth and constitutes the ``bouncing barrier''. It should be noted that relative velocities between two colliding grains were usually assumed to be monodisperse. Using relative velocity distributions accounting for stochastic motion, \citet{Garaud2013} showed that the low-velocity tail allowed collisional growth to larger sizes, thus overcoming both the bouncing and fragmentation barriers. A number of other barriers have also been studied, such as the charge barrier \citep{Okuzumi2009} and gravitational scattering of planetesimals \citep{Ida2008}.

Different solutions to the barriers of planet formation have been proposed. One example is the pathway to planetesimal formation via compression of fluffy aggregates developed by \citet{Kataoka2013}, which circumvents the radial-drift barrier. Another class of solutions are the so-called ``particle traps'', on which we will focus here. Aerodynamic drag ensures that dust particles follow the gas pressure gradient towards regions of high pressures, usually at the inner disk edge. If a well-defined pressure maximum exists elsewhere in the disk, grains will migrate towards it and accumulate there, thus escaping the radial-drift barrier. The higher concentration of dust in this ``trap'' will promote lower relative velocities and help grains to remain below the bouncing and fragmentation thresholds. Trapped grains will be able to grow more easily towards planetesimal sizes. Several types of particle traps have been studied: in anticyclonic vortices \citep{Barge1995,Regaly2012,Meheut2012}, at the snow line or at the dead zone \citep{Kretke2007,Dzyurkevich2010}, at planet gap edges \citep{Paardekooper2004,Fouchet2007,Fouchet2010,Ayliffe2012,Gonzalez2012}, or in any kind of ``pressure bump'' in the gas surface density \citep{Pinilla2012bump}. Particle traps have also been observed in protoplanetary disks, and regularly interpreted as due to vortices \citep[see e.g.][]{Casassus2013,vdM2013}.

The rest of this paper is devoted to particle traps at planet gap edges. In a previous work \citep{Fouchet2010}, we ran simulations to follow the dynamics of grains of a constant size in a viscous protoplanetary disk containing a planet using our 3D, two-fluid (gas+dust), SPH code \citep{BF2005}. We modeled a disk typical of Classical T~Tauri Stars (CTTS) of mass \mbox{$M_\mathrm{disk}=0.01\ M_\odot$} orbiting a 1~$M_\odot$ star and containing 1~\% of dust by mass, initially well mixed with the gas. We studied the evolution of the dust phase for three different grain sizes: $100~\mu$m, 1~mm and 1~cm, one at a time. We chose this size range because it encompasses the grains contributing the most to the disk emission at the ALMA wavelengths. We ran simulations for planets of 0.1, 0.5, 1 and 5~$M_\mathrm{J}$, on a circular orbit of radius 40~AU.

Previous simulations of planet gap formation in dusty gas disks were done in 2D, with a vertical integration of the disk structure. \citet{Paardekooper2004,Paardekooper2006} showed that planet gaps were more easily opened in the dust phase than in the gas. \citet{Rice2006} found the gap outer edge acted as a filter, holding back large dust grains while letting small grains through to the inner disk, a result later confirmed by \citet{Zhu2012}. With the third dimension, our resulting dust distributions accounted for dust settling and showed an even easier gap opening in the thinner dust disk \citep{Fouchet2010}. The gap's width and depth were larger for increasing planet mass and varied with grain size. In all cases, the gap created in the dust phase was deeper and wider than in the gas, as a result of the dust motion towards the gas pressure maxima at the gap edges. The dust density was largely enhanced there, demonstrating very efficient dust trapping for the considered grain sizes. A population of 1~cm grains was also found to be trapped in corotation with the most massive planet, on horseshoe orbits. \citet{Zhu2014} later found that gap formation in a 3D inviscid disk led to the Rossby-Wave Instability and the creation of vortices, with efficient trapping of grains across a large size range at gap edges and in vortices.

In a followup work, we used our resulting dust distributions to compute synthetic ALMA images of the simulated disks with a 1 and a 5~$M_\mathrm{J}$ planet \citep{Gonzalez2012}. We assessed the detectability of planet gaps for a variety of ALMA observing configurations (wavelength, integration time, angular resolution) and source properties (disk inclination, distance and declination). We found that, thanks to the dust accumulation at gap edges, gap detection is robust and that ALMA, when completed, should discover a large number of planet gaps in nearby star forming regions. Using 3D hydrodynamical and magneto-hydrodynamical simulations, assuming perfect mixing of non-growing grains with the gas, and radiative transfer on the resulting structures, \citet{Ruge2013} found similar conclusions for a wide range of disk masses and sizes around various central stars.

In this paper, we present new hydrodynamical simulations of the growth and fragmentation of grains. We detail the method in Sect.~\ref{Sec:Method}, show our results in Sect.~\ref{Sec:Results} and discuss them in Sect.~\ref{Sec:Discussion}. We investigate the observability of the resulting features in Sect.~\ref{Sec:Images} and conclude in Sect.~\ref{Sec:Concl}.

\section{Method}
\label{Sec:Method}

We now study the effect of grain growth and fragmentation on the dust dynamics in a disk containing a gap-opening planet. Our goal is to assess the efficiency of dust trapping at the gap edges to facilitate the growth of solids. We ran new hydrodynamical simulations with our SPH code using the same CTTS disk as in our previous work \citep[see Sect.~\ref{Sec:Intro}]{Fouchet2010}, with a 5~$M_\mathrm{J}$ planet on a 40~AU orbit. The code is extensively described in \citet{BF2005} and a discussion on the viscosity implementation can be found in \citet{Fouchet2007}. The SPH formalism has been shown to naturally reproduce the expected properties of turbulence \citep[see][]{Arena2013}. We use $\alpha_\mathrm{\scriptscriptstyle SPH}=0.1$ and $\beta_\mathrm{\scriptscriptstyle SPH}=0.5$, which corresponds to a uniform value of the \citet{SS1973} parameter $\alpha\sim10^{-2}$.

The gas disk, of mass $M_\mathrm{disk}=0.01\ M_\odot$, extends from 4 to 120~AU, has a flat surface density profile $\Sigma(R)=19.67$~kg\,m$^{-2}$ and is free to expand by viscous spreading. It is vertically isothermal and is initially setup to have a radial temperature profile $T(R)\propto R^{-1}$ with an aspect ratio of $H/R=0.05$ at the planet's location, translating in $T=15$~K at 40~AU.
We start by embedding the planet in the gas disk and we evolve the system for 8 planetary orbits to ensure an adequate relaxation of the gas phase \citep[see][]{BF2005,Fouchet2007,Fouchet2010}, before overlaying the dust phase with an initially uniform dust-to-gas ratio of 1\%. We assume that the dust grains, all with the same composition, are mostly made of water ice (the disk is almost entirely outside the snow line) and take their intrinsic density to be 1\,000~kg\,m$^{-3}$. Dust and gas interact via aerodynamic drag in the Epstein regime, which is appropriate for the local disk conditions. Particles would need to be larger than 10~m in the inner disk or 500~m in the outer disk to enter the Stokes regime, see Eq.~(44) and Fig.~5 of \citet{Laibe2012}. The corresponding Stokes number is
\begin{equation}
\mathrm{St}=\frac{\Omega_\mathrm{K}\rho_\mathrm{d}s}{\rho_\mathrm{g}c_\mathrm{s}},
\label{Eq:St}
\end{equation}
where $\Omega_\mathrm{K}$ is the Keplerian angular velocity, $\rho_\mathrm{d}$ the intrinsic dust density, $s$ the grain size, $\rho_\mathrm{g}$ the gas density and $c_\mathrm{s}$ its sound speed\footnote{The collisional velocities of gas molecules on grains involved in the drag force calculation in the Epstein regime is taken by some authors to be the gas sound speed, while others use the mean gas thermal velocity. We use the former definition, while the latter would have an additional factor of $\sqrt{\pi/8}$. Expressions of St for the Stokes regime can be derived from Appendix~C of \citet{Laibe2012}.}. We treat in a consistent manner the backreaction of dust on gas, as well as the gravitational force of the planet on grains \citep{Fouchet2010}. Both phases contain 200,000 SPH particles and all simulations were evolved for 100\,000~yr. We do not evolve the system beyond that time because our simulations do not include processes which occur on longer timescales (e.g. streaming instability, formation of planet embryos, disk photoevaporation).

Grain growth is implemented in our code as detailed in \citet{Laibe2008}, following the model of \citet{SV1997} who set up a ``sub-grid'' model for turbulence as a correlated noise to calculate the relative velocity $V_\mathrm{rel}$ between grains as a function of the velocity difference between dust and gas, the $\alpha$ parameter, $c_\mathrm{s}$ and St. Prescriptions differ whether only the vertical oscillations or also the epicyclic oscillations are considered \citep{Youdin2007}, see also \citet{Laibe2014} for a discussion on different expressions of $V_\mathrm{rel}$. Due to the nature of the SPH formalism, there is a natural spread in the velocity difference between gas and dust for a given grain size and disk location, thus producing a spread in the distribution of $V_\mathrm{rel}$ \citep[as described by][]{Windmark2012,Garaud2013}. Assuming compact icy grains that stick perfectly upon collision, with a locally monodisperse size distribution, \citet{SV1997} derived an analytical expression for the evolution of the grain size as a function of $V_\mathrm{rel}$, and thus of the local disk conditions.

We introduce a simple model for fragmentation by defining a velocity threshold, $V_\mathrm{frag}$, to which $V_\mathrm{rel}$ is compared. When $V_\mathrm{rel}<V_\mathrm{frag}$, grains grow and when $V_\mathrm{rel}>V_\mathrm{frag}$, they shatter, leading to a decrease of the size of the interacting SPH particles, modeled as a negative growth. The fragment size distribution is kept locally monodisperse to be consistent with the growth model, as well as to properly conserve physical quantities and avoid prohibitively large numbers of representative particles in the simulations. This means we only keep track of the largest fragment of the distribution: in the monodisperse case, the distribution relaxes locally to a size for which $V_\mathrm{rel}(s)\la V_\mathrm{frag}$. This approximation is relevant since we are interested in the disk's ability to form larger bodies. However, it cannot follow the smallest fragments who maintain a population of small grains throughout the disk. $V_\mathrm{frag}$ is a free parameter of our simulations and we assume it to be constant. This is clearly a simplification. The fragmentation velocity depends on the nature of the grains, such as composition and porosity, which we do not vary here. This simple, first order model is however very useful to understand the impact of the value of $V_\mathrm{frag}$. This approach is a common approximation in simulations of dust evolution in protoplanetary disks \citep[see, e.g.,][]{Birnstiel2010,Pinilla2012bump}, sometimes with a small transition interval near  $V_\mathrm{rel} \sim V_\mathrm{frag}$. Imperfect (or partial) sticking and bouncing are not taken into account. They would both contribute to lower the global growth efficiency: the outcome of a collision would be a smaller size than for perfect sticking or an unchanged size, respectively. Thresholds between different regimes are usually expressed in terms of $V_\mathrm{frag}$, and to first order the effect on the global growth efficiency amounts to varying $V_\mathrm{frag}$. We therefore chose to keep only one free parameter here.

In most studies of growth and fragmentation in protoplanetary disks, low values of $V_\mathrm{frag}$ were generally used: for example \citet{Birnstiel2010} took 1~m\,s$^{-1}$ for silicate aggregates and \citet{Pinilla2012bump} used 10~m\,s$^{-1}$ for icy material, following earlier results from laboratory experiments or theoretical work \citep[see, e.g.,][]{Blum2008}. More recently, larger values up to 20-30~m\,s$^{-1}$ have been considered \citep{Gonzalez2013,Meru2013,Pinilla2015}. Indeed, according to recent studies, grain growth seems to be able to proceed at higher relative velocities. First, grain porosity seems to be an important factor. \citet{Wada2009} ran $N$-body simulations of collisions of porous aggregates and determined that \mbox{$V_\mathrm{frag}\sim60$~m\,s$^{-1}$} for icy aggregates but only $\sim6$~m\,s$^{-1}$ for silicates. \citet{Meru2013} studied dust collisions with SPH simulations and found that $V_\mathrm{frag}$ can exceed 27~m\,s$^{-1}$ for cm-sized porous silicate aggregates of different masses. Mass transfer in high mass ratio collisions is a second factor facilitating the growth of solids. In laboratory experiments of such collisions, \citet{Teiser2009} showed that $V_\mathrm{frag}\sim60$~m\,s$^{-1}$ for silicate aggregates. With their $N$-body simulations, \citet{Wada2013} obtained $V_\mathrm{frag}\sim80$~m\,s$^{-1}$ for icy aggregates but $\sim8$~m\,s$^{-1}$ for silicates. More recently, the same group used updated values of the surface energies \citep{Yamamoto2014} and found $V_\mathrm{frag}\sim30-40$~m\,s$^{-1}$ for silicate aggregates and even as high as 100~m\,s$^{-1}$ for iron, reconciling their numerical results with those from experiments. We chose to run simulations for fragmentation threshold values of $V_\mathrm{frag}=10$, 15, 20 and 25~m\,s$^{-1}$. Those values adequately sample the growth behavior between cases where fragmentation is too efficient to allow any significant growth and where growth is almost unhindered by rare fragmentation. Extreme cases on both sides are the non-growing grains (equivalent to $V_\mathrm{frag}=0$) we studied in \citet{Fouchet2010} and the pure growth case (equivalent to $V_\mathrm{frag}=+\infty$) treated in this paper.

We start from an initially uniform grain size $s_0=10\ \mu$m, and also take this value to be the smallest grain size that results from fragmentation. Test simulations with smaller values of $s_0$ showed that very small grains grow fast and quickly forget their initial size \citep[see also][]{Laibe2008}. We chose this larger value to shorten computation times. Note that 10~$\mu$m grains never fragment since they have $V_\mathrm{rel}<10$~m\,s$^{-1}$, our lower value of $V_\mathrm{frag}$, over the whole disk and smaller grains would have even smaller $V_\mathrm{rel}$. Fragmentation therefore does not affect this conclusion.

\section{Results}
\label{Sec:Results}

\begin{figure}
\centering
\resizebox{\hsize}{!}{
\includegraphics{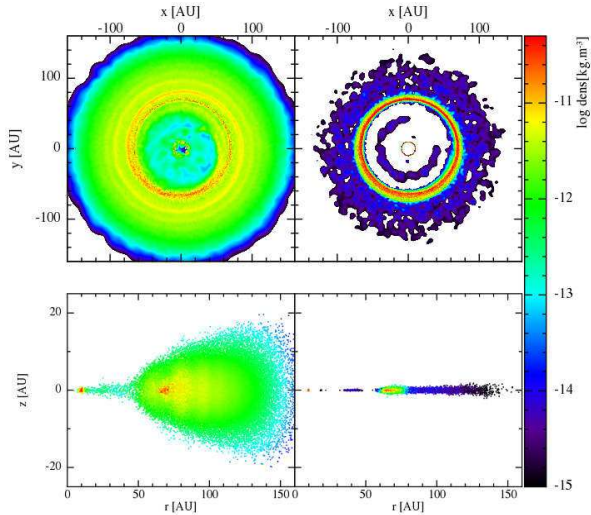}
}
\caption{Volume density maps in midplane (top) and meridian plane (bottom) cuts of the disk for the gas (left) and dust (right) after 50\,000~yr for the simulation with pure grain growth.}
\label{Fig:xy_rz}
\end{figure}

Figure~\ref{Fig:xy_rz} shows volume density maps of the gas and dust phases after 50\,000~yr of evolution of the simulation with pure growth. The planet has opened a deep, but not empty, gap in the gas disk, as well as launched spiral waves in the outer disk (i.e. the disk exterior to the gap). In the dust disk, the gap is cleared, except for a population of dust grains on horseshoe orbits in corotation with the planet. Two narrow rings of dense dust are present on both sides of the gap, showing the efficient dust accumulation. There, the dust-to-gas ratio exceeds unity: it is $\sim5$ in the outer ring and $\sim50$ in the inner ring. The dust disk has a smaller outer radius than the gas disk, because of a moderate radial migration of the (still small) grains in the very outer disk regions. The dust disk is also thin due to an efficient dust settling. Overall, the disk appearance is largely similar to what was seen for non-growing grains \citep[see][]{Fouchet2010}, except in the inner disk (i.e. the disk interior to the gap) where growth helps grains overcome the radial-drift barrier and pile up, consistent with the \citet{Laibe2014} mechanism. The behavior of the gas disk is similar in all simulations, we will now focus on the dust phase.

\begin{sidewaysfigure}
\centering
\resizebox{\hsize}{!}{
\includegraphics{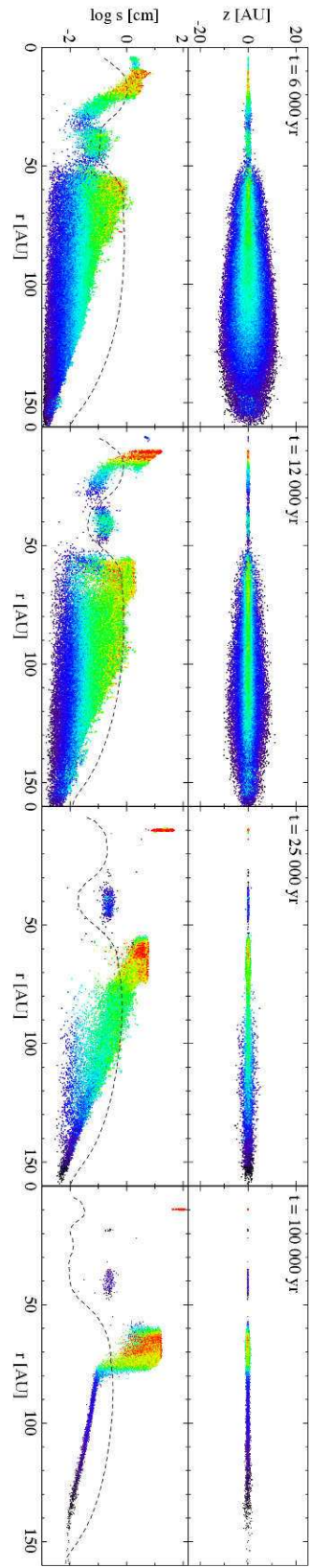}
}
\caption{Time evolution of the dust phase in the simulation with pure growth. \textit{Top:} Meridian plane cut of the dust distribution. \textit{Bottom:} Radial grain size distribution. \textit{From left to right:} Snapshots at $t=6\,000$, 12\,000, 25\,000 and 100\,000~yr. The color represents the volume density and is the same as in Fig.~\ref{Fig:xy_rz}. The dashed line marks the grain size for which $\mathrm{St}=1$ in the disk midplane.}
\label{Fig:Vinf}
\end{sidewaysfigure}

The evolution of the dust population is best seen is Fig.~\ref{Fig:Vinf}, which shows its distribution in the meridian plane together with the radial distribution of grain sizes at $t=6\,000$, 12\,000, 25\,000 and 100\,000~yr ($\sim25$, 50, 100 and 400 planetary orbits, respectively). The dashed curves show the grain sizes for which $\mathrm{St}=1$ in the midplane (they are proportional to the gas surface density --- see Eq.~(\ref{Eq:s_opt}) --- and therefore also trace the evolution of the gas phase). Similarly to what was found by \citet{Laibe2008} in a CTTS disk without planets, particles typically grow as they settle to the midplane, then migrate rapidly when they reach a size for which $\mathrm{St}=1$ while experiencing little growth. Grains initially close to the disk inner edge are thus lost to the star, as is the case for the moderately dense clump of centimeter-sized particles seen interior to 10~AU in the 6\,000~yr snapshot. A little further out we also see grains accumulate interior to the planet gap. Here their density is high and they can grow efficiently. A similar behavior is seen in the outer disk, where grains migrate inwards to the outer edge of the planet gap where they pile up and grow. Particles in the gap drift towards one of the gap edges. The growth timescale varies as $\Omega_\mathrm{K}^{-1}\propto r^{3/2}$ \citep{Laibe2008} explaining the smaller grain sizes in the farthest regions. At 12\,000~yr, the detached group of particles at the disk inner edge contains the last grains to be lost to the star. Just outside of 10~AU, particles have outgrown the $\mathrm{St}=1$ size ($\sim1$~cm) and piled up, are now decoupled from the gas and grow without migrating. The gap continues to empty, except for grains close to the planet's orbit. They have outgrown the $\mathrm{St}=1$ size, which rapidly decreased together with the gas surface density (Eq.~(\ref{Eq:s_opt})), and are now in corotation with the planet \citep[see][for a discussion on particle trajectories in the horseshoe region]{Fouchet2010}. At 25\,000~yr, all grains in the inner disk are concentrated in a narrow, dense ring where they continue to grow. The outer gap edge has also become very dense. The evolution is subsequently very slow and finally, at 100\,000~yr, the gap only contains the mm-sized particles that are trapped in corotation with the planet. Their density is very low and they behave as isolated solids and no longer grow. Large grains are present on both sides of the gap, showing the efficiency of the pile-up in the inner disk and the particle trap at the outer gap edge in promoting the formation of solids larger than the cm sizes. One might wonder whether the dust accumulation in the inner disk may dissipate on longer timescales. The gas disk interior to the gap will indeed be accreted onto the star on viscous timescales, however, once the grains have outgrown the $\mathrm{St}=1$ size and decoupled from the gas, they will remain in their dense ring and continue to grow, whether gas is present or not (one can notice that after 100\,000~yr the gas surface density has already decreased by a factor of a few tens compared to its initial value).

\begin{sidewaysfigure*}
\centering
\resizebox{\hsize}{!}{
\includegraphics{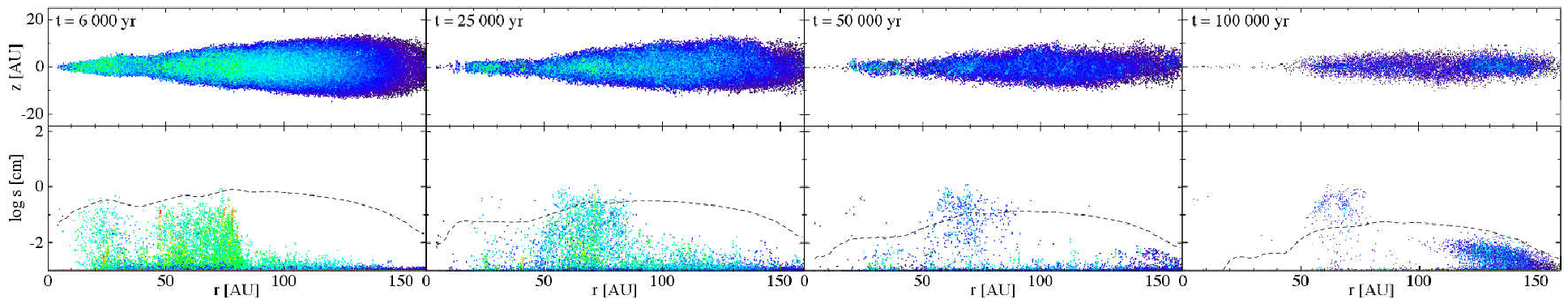}
}
\resizebox{\hsize}{!}{
\includegraphics{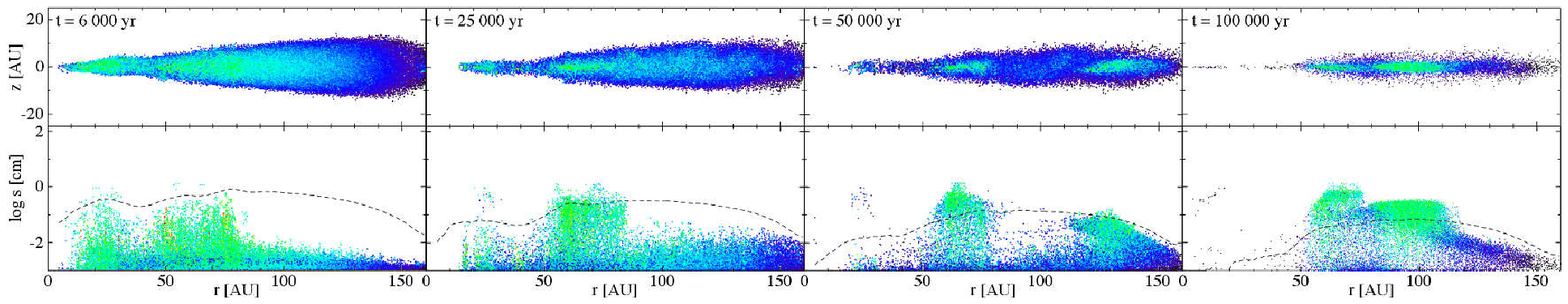}
}
\resizebox{\hsize}{!}{
\includegraphics{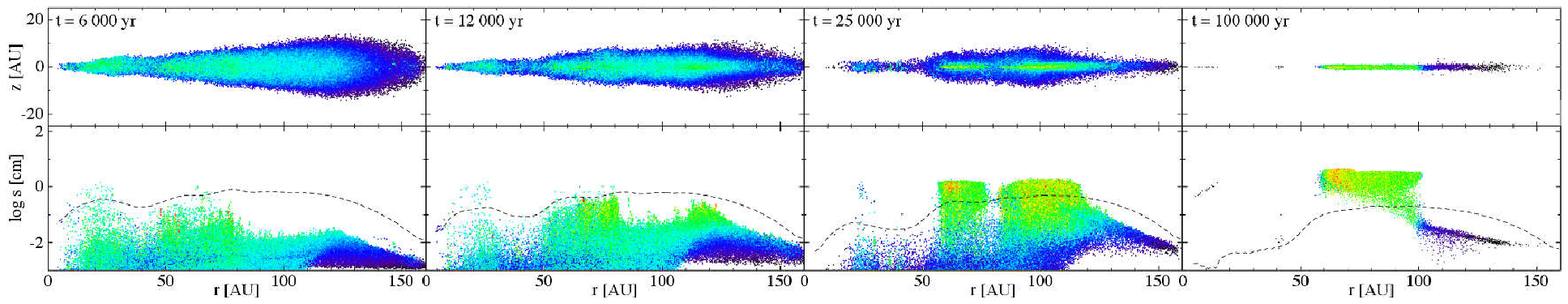}
}
\resizebox{\hsize}{!}{
\includegraphics{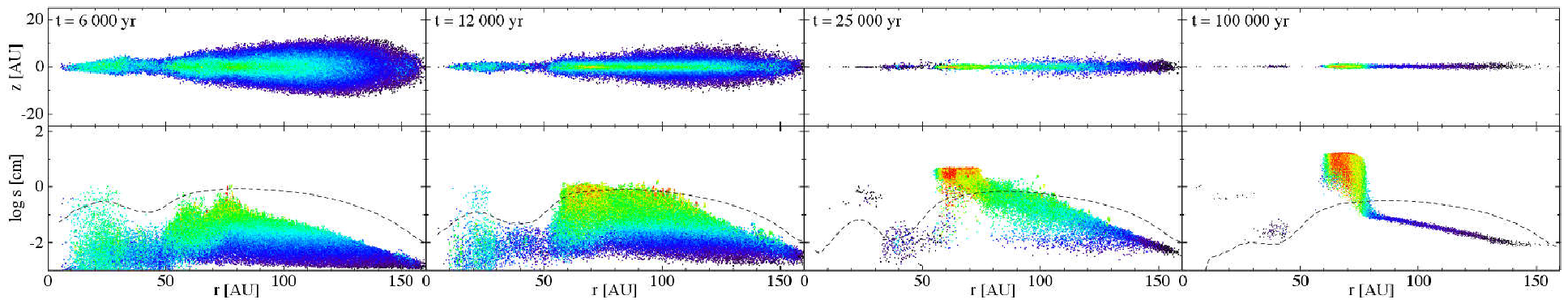}
}
\caption{Same as Fig.~\ref{Fig:Vinf} for the simulations with fragmentation for $V_\mathrm{frag}=10$, 15, 20 and 25~m\,s$^{-1}$, from top to bottom. Note that snapshots are shown at different times for each case, to better show their evolution.}
\label{Fig:V10-25}
\end{sidewaysfigure*}

Adding fragmentation has a dramatic effect on the dust evolution, especially for smaller values of $V_\mathrm{frag}$, as can be seen in Fig.~\ref{Fig:V10-25}, which shows the same plots as Fig.~\ref{Fig:Vinf} for $V_\mathrm{frag}=10$, 15, 20 and 25~m\,s$^{-1}$. For the lower two values of the fragmentation threshold (top two rows of Fig.~\ref{Fig:V10-25}), grains are not able to grow large enough in the inner disk to overcome the radial-drift barrier. The inner disk is thus seen to progressively be lost to the star.
For $V_\mathrm{frag}=10$~m\,s$^{-1}$, fragmentation keeps the grains small and prevents the majority of them from decoupling from the gas. Contrary to what occurs for larger grains which produce sharper and deeper gaps, no clear depletion in the dust is seen in the vicinity of the planet. Grains in the outer disk therefore follow the gas through the gap and migrate into the inner disk where their inward drift continues. This is similar to the filtering of small grains by the gap seen in the non-growing case \citep{Rice2006,Zhu2012}. The dust disk slowly drains and its density after 50\,000~yr is very low. At the disk outer edge, the relative velocities between grains \citep[which decrease towards larger radii, see][]{SV1997} are low enough to remain below the fragmentation threshold and allow their growth. However, this growth is very slow, its timescale being longer at larger radii, and at 100,000~yr, grains have only attained $\sim100~\mu$m.
For $V_\mathrm{frag}=15$~m\,s$^{-1}$, the planet gap is shallow, but the density difference is large enough to trap grains at the outer gap edge, where they start to overcome the $\mathrm{St}=1$ size after 25\,000~yr and grow slowly. They exceed centimeter sizes at 50\,000~yr. In the outer disk regions, grains stay below the fragmentation threshold and growth, contrary to the intermediate regions which remain dominated by fragmentation. At 100,000~yr, grains in that outer population have grown to a few millimeters and migrated to form a denser annulus closer to the gap outer edge. For both $V_\mathrm{frag}=10$ and 15~m\,s$^{-1}$, grains do not decouple enough from the gas in the gap to be trapped in corotation with the planet.

The situation improves for $V_\mathrm{frag}=20$ and 25~m\,s$^{-1}$ (bottom two rows of Fig.~\ref{Fig:V10-25}). The higher values of the threshold help to retain more grains in the inner disk and a fraction of them overcomes the radial-drift barrier. A small population of grains survive in the inner disk after 25\,000~yr, but do not grow much larger than cm sizes due to a limited reservoir from which to collect mass. For $V_\mathrm{frag}=20$~m\,s$^{-1}$, growth is more efficient at the outer gap edge as well as in the disk outer regions (fragmentation is still significant in the intermediate regions). At 100\,000~yr, both populations have merged to form an extended and more settled ring of centimeter-sized grains. For $V_\mathrm{frag}=25$~m\,s$^{-1}$, some particles manage to grow in the gap to a few hundred micrometers before being trapped in the horseshoe region. Because of less efficient growth, the grains sizes are smaller when they decouple from the gas than in the pure growth case. The outer disk is almost unaffected by fragmentation: it is very similar to the case with pure growth, with very efficient growth past cm sizes in a dense ring at the outer gap edge.

\section{Discussion}
\label{Sec:Discussion}

The accumulation of grains on either side of the planet gap lowers they relative velocities and increases their density, thereby facilitating their growth. When fragmentation is not taken into account, this growth is very efficient and can lead to the formation of solids larger than the decimeter, and even reaching meter sizes in the pile-up of the inner disk, in as little as 50\,000~yr. The outer edge of the gap created by the planet embedded in the disk appears as a potential site for the formation of additional planets. Indeed, even a forming giant planet can carve a shallow gap  \citep[as was seen for a planet mass of 0.1~$M_\mathrm{J}\sim30\ M_\oplus$ by][]{Fouchet2010}. The pressure maximum at the outer edge of a planet gap traps dust grains and can start the growth process before the first planet is fully formed. This mechanism could lead to a system of giant planets.

Fragmentation, however, makes planetesimal formation more difficult. When included in the simulations, it causes different growth behaviors depending on the distance from the star. This is essentially due to the radial dependence of the relative velocity of grains. To first order, the relative velocities are proportional to the gas sound speed, which varies as $R^{-1/2}$ in our disk, with a dispersion due to turbulence \citep[see][]{SV1997}. In the inner disk, $V_\mathrm{rel}$ is large and grain growth is quickly limited by the fragmentation threshold $V_\mathrm{frag}$. In the outer disk, the lower values of $V_\mathrm{rel}$ allow growth to proceed more easily, as was seen for all values of $V_\mathrm{frag}$ considered here at the outer disk edge. Obviously, large values of $V_\mathrm{frag}$ allow particles to reach larger sizes and survive more easily in the disk. On the contrary, for a low $V_\mathrm{frag}$ no significant growth past the radial-drift barrier is observed. The role of fragmentation as one of the barriers of planet formation is clearly illustrated here.

Varying the initial conditions can have an influence on the results. The main parameter governing grain dynamics is the optimal size of migration, for which $\mathrm{St}=1$, that can be obtained from Eq.~(\ref{Eq:St}). Its value in the midplane is
\begin{equation}
s(\mathrm{St}=1)=\frac{\Sigma_\mathrm{g}}{\sqrt{2\pi}\,\rho_\mathrm{d}}
\label{Eq:s_opt}
\end{equation}
\citep{Fouchet2010} and depends only on the gas surface density, scaling linearly with the gas disk mass. When fragmentation dominates, grains always stay below $\mathrm{St}=1$. Otherwise, small grains grow fast until they reach the optimal size and changing its value only translates the radial size distribution \citep[the ``migration plateau'', see][]{Laibe2014} and thus the larger sizes grains can reach after decoupling. A more massive disk produces larger grains but does not affect the location of particle traps.
Changing the disk temperature has a more moderate effect. Indeed, the relative velocities between grains scale as the gas sound speed, and thus as the square root of the temperature. Producing similar results with a warmer or colder disk only requires a small increase or decrease of $V_\mathrm{frag}$.
We use an initially uniform dust-to-gas ratio of 1\%, a commonly used number taken from the interstellar medium value. However, \citet{Williams2014} measured it to be 5\% on average (with a large spread). Higher values of the dust-to-gas ratio reduce the migration efficiency \citep{NSH1986}, as well as accelerate grain growth and thus increase their pile-up \citep{Brauer2008,Laibe2014,Pinte2014}. This therefore favors grain survival and planetesimal formation.

We do not vary the planet mass in this paper. As we found (for non-growing grains) in \citet{Fouchet2010}, as long as the planet is massive enough to open even a partial gap, it will trap grains. More massive planets produce a steeper pressure gradient, which makes it easier for grains to accumulate and grow, requiring a small value of $V_\mathrm{frag}$. In this type~II migration regime, allowing the planet to migrate is not expected to have a significant impact on the overall results since the timescale for dust evolution is much shorter than the planet migration timescale.

Similar studies of dust traps and pile ups in disks with planets have been conducted by \citet{Pinilla2012gap} and \citet{Pinilla2015} for a range of planet and disk configurations which vary parameters such as the stellar mass, the planet mass and semi-major axis, the disk turbulence (via $\alpha$) and disk mass, and the dust fragmentation threshold ($V_\mathrm{frag}$). In \citet{Pinilla2012gap} they studied the dust dynamics in a disk with one planet, while in \citet{Pinilla2015} the effects of two embedded planets was investigated. Both studies used 2D hydrodynamical simulations, for which the resulting disk structure was used as a static gas backdrop to follow the dust evolution with a 1D coagulation/fragmentation model, without backreaction. In order to compare the results of our simulations with these works, there are two main quantities that need to be determined: (i) the gap opening criterion for viscous disks \citep{Lin1979,Bryden1999}, which depends on the ratio of the planet to stellar mass, $M_\mathrm{p}/M_\star$, as well as the disk aspect ratio at the location of the planet, $H/r_\mathrm{p}$, and the turbulence parameter, $\alpha$; and (ii) the ratio of the gas turbulent velocity to the dust fragmentation threshold, $V_\mathrm{turb}/V_\mathrm{frag}$ (since the turbulent gas velocity sets the maximum value of the relative dust velocity, $V_\mathrm{rel}$, and grains will grow when $V_\mathrm{rel}/V_\mathrm{frag} < 1$).  $V_\mathrm{turb}$ is proportional to $c_s\sqrt{\alpha}$ and a factor that depends on the grain size.

In \citet{Pinilla2012gap} all the disks they studied open strong gaps. Unfortunately we cannot calculate $V_\mathrm{turb}/V_\mathrm{frag}$ for their disks as they do not provide details of the disk temperature which is needed to determine $c_s$ and $H/r_p$. However, with high $\alpha/V_\mathrm{frag}$ ($\alpha=10^{-2}$ and $V_\mathrm{frag}=10$~m\,s$^{-1}$), they found no dust trapping for a 1~$M_\mathrm{J}$ planet and the disk emptied of dust. When $\alpha$ was decreased to $10^{-3}$, they found that dust could be trapped at the outer gap edge and grow, since a lower $\alpha$ leads to smaller relative dust velocities which can more easily remain below $V_\mathrm{frag}$. With a 9~$M_\mathrm{J}$ planet, they found dust trapping for both values of $\alpha$, which is expected to be easier as the planet mass increases. In \citet{Pinilla2015} they again found dust trapping and growth (at the outer edge of both planet gaps).  All their simulations are in the strong gap opening regime, and $V_\mathrm{turb}/V_\mathrm{frag}$ ensures dust trapping and growth, as in our simulations.

Our simulations show a very efficient dust trapping and growth at the outer gap edge for disk plus planet configurations that are in the strong gap opening regime and for favourable $V_\mathrm{turb}/V_\mathrm{frag}$ (in our case when $V_\mathrm{frag}=25$~m\,s$^{-1}$). Furthermore, as was shown by \citet{Fouchet2007,Fouchet2010}, grain accumulation at gap edges is stronger in the settled dust layer of a 3D disk than in a vertically integrated 2D disk. It should also be noted that it is easier to trap dust in planet gaps around a low mass star, where gap opening is easier and the disk is cooler and hence gas turbulent velocities are lower.

Simulations show that grain growth to centimeter sizes requires fragmentation thresholds larger than 10 or 20~m\,s$^{-1}$, depending on the disk turbulence. Although those values seemed too high according to earlier laboratory experiments \citep{Blum2008}, recent work shows that values of $V_\mathrm{frag}$ of several tens of m\,s$^{-1}$ are now realistic when considering collisions of porous aggregates or mass transfer in high mass ratio collisions, or a combination of both (see Sect.~\ref{Sec:Method}). Even if these two cases represent only a fraction of all grain collisions in protoplanetary disks, they may be enough to ensure that at least some of the dust population manages to grow to eventually form planetesimals.

\section{Synthetic images}
\label{Sec:Images}

\begin{sidewaysfigure*}
\centering
\resizebox{\hsize}{!}{
\includegraphics{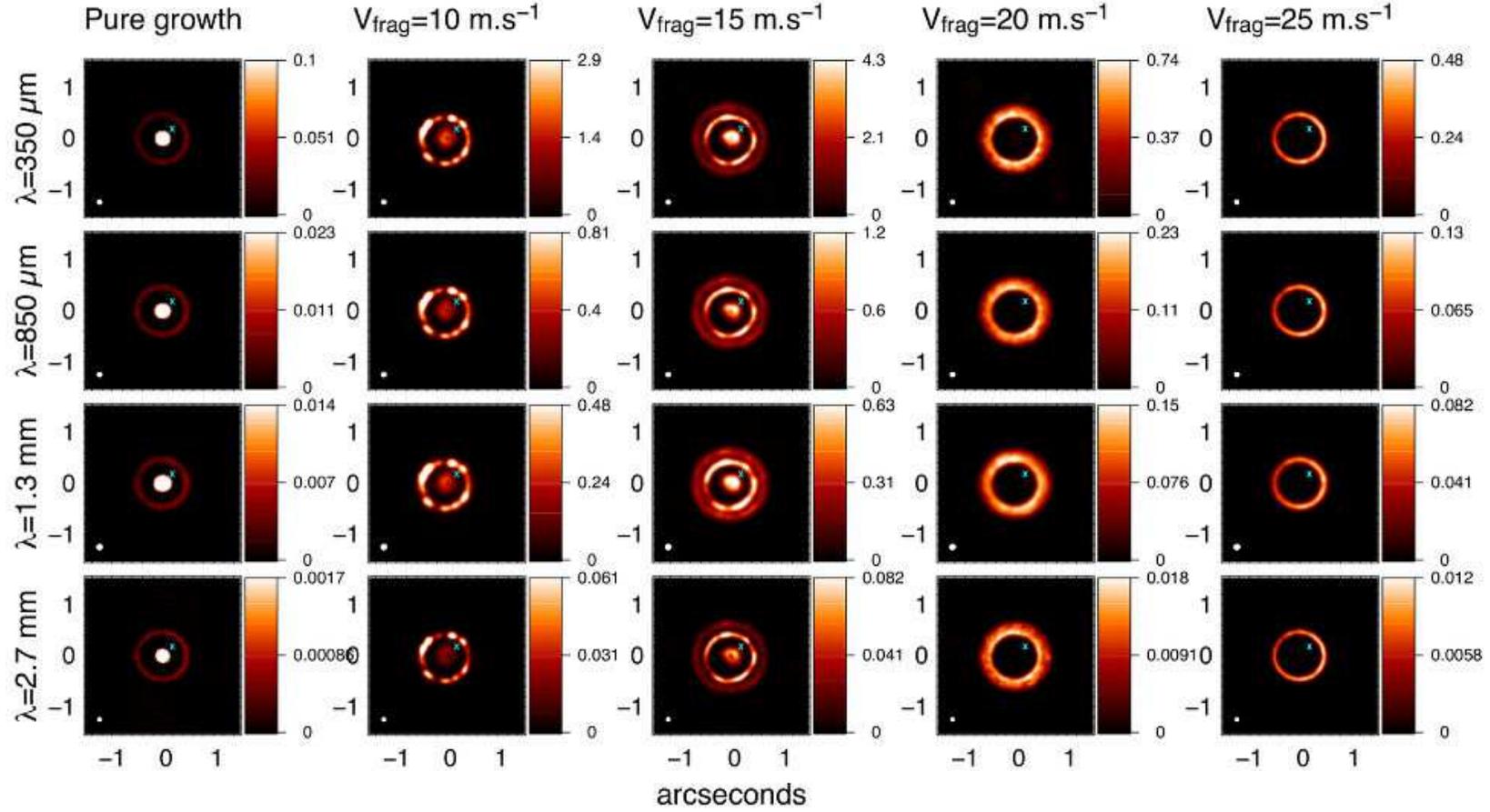}
}
\caption{Simulated ALMA observations of a disk viewed face-on at $d=140$~pc and $\delta=-23^\circ$ for an integration time of 1~h and angular resolution of $0.1''$. \textit{From left to right:} Simulations with pure growth and with fragmentation for $V_\mathrm{frag}=10$, 15, 20 and 25~m\,s$^{-1}$. \textit{From top to bottom:} $\lambda=350\ \mu$m, 850~$\mu$m, 1.3~mm and 2.7~mm. The scale on each image is in arcseconds, with the beam size represented at its bottom left corner, and the colorbar gives the flux in mJy/beam. The planet's position is marked with a cross.}
\label{Fig:ALMA}
\end{sidewaysfigure*}

In order to determine the impact of growth and fragmentation on the planet gap detectability with ALMA and to assess whether the fragmentation threshold can be constrained by observations, we computed synthetic images from the resulting disk structure of each of our simulations after 100,000~yr. We first used the 3D Monte Carlo continuum radiative transfer code \textsf{MCFOST} \citep{Pinte2006,Pinte2009} to produce raw intensity maps from the dust distributions. Dust grains are taken to be homogeneous spheres and composed of astronomical silicates \citep{Weingartner2001}, their optical properties are computed according to the Mie theory. These maps were then passed to the CASA\footnote{\texttt{http://casa.nrao.edu}} simulator for ALMA, to obtain synthetic images for a given observing configuration (wavelength, angular resolution, integration time). The procedure is described in detail in \citet{Gonzalez2012}.

To allow for a better comparison with the images produced in the case of non-growing grains, we chose the standard disk parameters of \citet{Gonzalez2012}: nearly face-on orientation, a distance $d=140$~pc and a declination $\delta=-23^\circ$ (for which the source passes through the zenith at the ALMA site) and their optimal observing parameters for gap detection: integration time $t=1$~h and angular resolution $\theta=0.1''$ for 4 different wavelengths: $350~\mu$m, $850~\mu$m, 1.3~mm and 2.7~mm. The images for pure growth and the four different fragmentation thresholds are shown in Fig.~\ref{Fig:ALMA}.

The most striking feature of these images is that the gap is not always visible. One has to keep in mind that the disk appearance at a given wavelength results from a combination of the dust density and the grain size. In the case of non-growing grains \citep{Gonzalez2012}, computations chose grain sizes that contribute the most to the ALMA wavelengths and the planet gap was prominent in all cases. For growing and fragmenting grains, their sizes evolve and may end up outside the optimal ALMA range ($s\sim20~\mu$m -- 1~mm for $\lambda=350~\mu$m to 2.7~mm) for the bulk of the dust population. This is the case when grains can efficiently reach large sizes: in the pure growth case or when $V_\mathrm{frag}\ge20$~m\,s$^{-1}$. Only a small fraction of grains have the appropriate sizes in a thin annulus in the outer disk, producing a faint (note the colorbar scales in Fig.~\ref{Fig:ALMA}) ring in the images. For the pure growth case, the density is so high in the inner ring that, even though the grain sizes are outside the optimal range, the tail end of their emission is still large enough to be detectable. It is only for $V_\mathrm{frag}=10$ and 15~m\,s$^{-1}$ that there is a sufficiently large population of grains of the right size on both sides of the gap for them to appear bright, thus producing a prominent gap. Large fragmentation thresholds, producing large solids and reducing the small grain reservoir, are therefore favorable for planet formation but not for disk observations and in particular for planet gap detection.

Note that for $V_\mathrm{frag}=15$~m\,s$^{-1}$, a second ring is visible exterior to the outer gap edge. It shows the population of grains that grew slowly in the outer disk, separated from the gap edge by a region of efficient fragmentation. The occurrence of such multiple rings, as recently observed in the HL Tau disk \citep{Brogan2015}, will be examined in a forthcoming paper. For $V_\mathrm{frag}=10$~m\,s$^{-1}$, growth at the outer disk edge is too slow to produce grains large enough to be visible at the ALMA wavelengths and only the bright ring at the outer gap edge stands out (bright spots along the ring are due to the rendering of a non-uniform azimuthal distribution of SPH particles).

Finally, the differences between each case are not significant enough to unambiguously discriminate one from the other. Different values of $V_\mathrm{frag}$ can produce similar structures at different evolutionary times, for example the outer population of growing grains causing the exterior ring seen with $V_\mathrm{frag}=15$~m\,s$^{-1}$ at 100,000~yr is similar to that seen for $V_\mathrm{frag}=20$~m\,s$^{-1}$ at 25,000~yr, before it merged with the dust population at the outer gap edge (Fig.~\ref{Fig:V10-25}, third row). The different disk intensities may be interpreted as different total dust masses, and the location of the observed ring by different planet locations. The problem is too degenerate to reach firm conclusions and determining the fragmentation threshold therefore seems very difficult from submillimeter observations alone. Data from other wavelength ranges, tracing different grain sizes, can bring additional constraints.

\section{Conclusion}
\label{Sec:Concl}

We have run 3D hydrodynamical simulations of the evolution of dust grains in a protoplanetary disk containing a planet in order to study the effect of growth and fragmentation on the formation of large solids at ``particle traps'' located at the edges of planet gaps. Such traps have been proposed as a solution to overcome the barriers of planet formation and allow the formation of planetesimals. We implemented a simple model for fragmentation, using a locally monodisperse size distribution, and found that it strongly limits the growth of dust grains even in the presence of dust traps and, in combination with radial drift, contributes to the loss of the inner disk. Only large values of the relative velocity threshold for fragmentation, taken as a free parameter ($V_\mathrm{frag}\ge20$~m\,s$^{-1}$) allow grains to grow above centimeter sizes after 100,000~yr. The value of the fragmentation threshold was usually thought to be only a few m\,s$^{-1}$, however recent experimental and numerical work has shown that coagulation at large velocities is possible under certain conditions (porous grains, or high mass ratio collisions), opening a possible path to planetesimal formation.

We produced synthetic ALMA images from our simulated disks and found that gap detection is made more difficult by large values of the fragmentation threshold. On the contrary, for intermediate values of $V_\mathrm{frag}$, a population of slowly growing grains in the outer disk can produce a second ring exterior to the outer gap edge in the ALMA images (after 100,000~yr for $V_\mathrm{frag}=15$~m\,s$^{-1}$ or earlier for larger $V_\mathrm{frag}$), reminiscent of the multiple rings observed in the HL~Tau disk. However, discriminating between different values of $V_\mathrm{frag}$ from submillimeter images seems impractical without additional constraints.

In a forthcoming study, we will improve our growth and fragmentation model by taking into account the grain porosity and its evolution during collisions.

\section*{Acknowledgments}

We thank the two anonymous referees for constructive feedback which helped us improve this manuscript.
This research was partially supported by the Programme National de
Physique Stellaire and the Programme National de Plan\'etologie of CNRS/INSU,
France, and the Agence Nationale de la Recherche (ANR) of France through
contract ANR-07-BLAN-0221.
J.-F. Gonzalez's research was conducted within the Lyon Institute of Origins
under grant ANR-10-LABX-66.
G. Laibe is grateful for funding from the European Research Council for the FP7 ERC advanced grant project ECOGAL.
C. Pinte acknowledges funding from the European Commission's FP7
(contract PERG06-GA-2009-256513) and ANR (contract ANR-2010-JCJC-0504-01).
Computations were performed at the Service Commun de Calcul Intensif de
l'Observatoire de Grenoble (SCCI).
Figures~\ref{Fig:xy_rz} to \ref{Fig:V10-25} were made with SPLASH \citep{Price2007}.

\section*{References}

\bibliographystyle{elsarticle-harv} 
\bibliography{gonzalez}

\end{document}